\documentstyle[twoside,fleqn,espcrc2]{article}
\def\mev{\,{\rm Me\kern-0.1em V}}
\def\gev{\,{\rm Ge\kern-0.1em V}}

\title{Resolving Exceptional Configurations} 

\author{
 W.~Bardeen\address{Fermilab, P.O. Box 500, Batavia, IL 60510},%
 A.~Duncan\address{Dept. of Physics and Astronomy,University of Pittsburgh,
 Pittsburgh, PA 15260},%
 E.~Eichten\thanks{Presenter}$^{\rm a}$,%
 G.~Hockney\address{Jet Propulsion Laboratory, 
 California Institute of Technology, Pasadena, CA 91109},%
 and 
 H.~Thacker\address{Dept. of Physics, University of Virginia, 
 Charlottesville, VA 22901}
}
\begin{document}
\begin{abstract}
In lattice QCD with Wilson fermions, exceptional configurations
arise in the quenched approximation at small quark mass.
The origin of these large previously uncontrolled lattice artifacts 
is identified. A simple well-defined procedure (MQA) is presented 
which removes the artifacts while preserving the correct continuum limit.
\end{abstract}
\maketitle
%
%
%
\section{Introduction}
%

Quenched lattice QCD studies using Wilson-Dirac fermions have
encountered large statistical errors in calculations involving
very light quarks. A small subset 
of the gauge configurations seem to play a dominant role 
in the final results. These ``exceptional configurations'' 
have limited studies to quark masses much larger than the up and down quark masses.
The large fluctuations resulting from these exceptional configurations
appear to be related to the structure of the small eigenvalues of the
Wilson-Dirac operator~\cite{Mutter86,Itoh87,Iwasaki89}. 

As I will explain, exceptional configurations
arise from a specific lattice artifact associated with
the Wilson-Dirac formulation.  
This artifact is not removed by using the standard
Sheikholeslami-Wohlert(Clover) $O(a)$ ``improved'' action~\cite{SWaction}. 
However, introducing a modified quenched approximation (MQA), 
removes the dominant part of this artifact and  
eliminates the exceptional configurations, thereby
greatly reducing the noise associated with light quarks.
The detailed studies that form the basis of this report 
are presented elsewhere~\cite{QCD4d97,QED2d97}.

%
\section{Eigenvalue Spectrum and Zero Modes}
%

The fermion propagator in a background gauge field, 
$A_{\mu}(x)$, may be written in terms of a sum over the eigenvalues 
of the Dirac operator,
\begin{equation}
{\cal D}f_i = \gamma*D f_i =\lambda_i f_i
\end{equation}    
and 	
\begin{equation}
S(x,y;\{A\}) = \sum_{eigenvalues} 
\frac{f_{i}(x;A)\bar{g}_{i}(y,A)}{(\lambda_i+m_0)}
\end{equation}       
where $f_i(x,A)$ and $\bar{g}_i(x,A)$ are the corresponding 
left and right eigenfunctions.   The mass dependence of the propagator is 
determined by the nature of the eigenvalue spectrum of the 
Dirac operator.   In the continuum, the euclidean Dirac 
operator is skewhermitian and its eigenvalues 
are purely imaginary or zero.   Hence, the fermion 
propagators only have singularities when  $m_0=0$. 
The behavior of the eigenvalue 
spectrum near zero determines the nature of dynamical 
chiral symmetry breaking \cite{Banks80}.   The zero eigenvalues, or zero 
modes, are related to topological fluctuations of the 
background gauge field by the index theorem associated with the 
chiral gauge anomaly \cite{vanBaal97}.    

The lattice formulation of Wilson-Dirac
fer\-mions qualitatively modifies the nature of the eigenvalue spectrum.   
The Wilson-Dirac operator is usually written as 
\begin{eqnarray}
\label{eq:Mdef}
{\cal D} & \equiv & D-rW \\
\label{eq:Ddef}
D_{a\alpha\vec{m},b\beta\vec{n}} & = & \frac{1}{2}(\gamma_{\mu})_{ab}
U_{\alpha\beta}(\vec{m}\mu)\delta_{\vec{n},\vec{m}+\hat{\mu}} \\
& & ~~-\frac{1}{2}(\gamma_{\mu})_{ab}U^{\dagger}_{\alpha\beta}
(\vec{n}\mu)\delta_{\vec{n},\vec{m}-\hat{\mu}} \nonumber \\
\label{eq:Wdef}
W_{a\alpha\vec{m},b\beta\vec{n}} & = & \frac{1}{2}\delta_{ab}
(U_{\alpha\beta}(\vec{m}\mu)\delta_{\vec{n},\vec{m}+\hat{\mu}} \\
 & & ~~~~~+U^{\dagger}_{\alpha\beta}(\vec{n}\mu)\delta_{\vec{n},\vec{m}-\hat{\mu}} ) \nonumber
\end{eqnarray}
where $U(\vec{n}\mu)$ are the link matrices associated with the lattice 
gauge fields, and the parameter $r$ is usually taken to be $1$.   
The Wilson-Dirac operator is neither skewhermitian (unless 
$r=0$) nor hermitian.   The Wilson term explicitly breaks chiral 
symmetry and lifts the doubling degeneracy of the pure lattice 
Dirac action.   As a result the eigenvalue spectrum of the 
Wilson-Dirac operator is no longer purely imaginary but fills 
a region of the complex plane. 
The discrete symmetries of the 
Wilson-Dirac operator imply that the eigenvalues appear in 
complex conjugate pairs, $(\lambda,{\lambda}^*)$ and obey 
reflection symmetries, $(\lambda,-\lambda)$.   In addition, 
there can be precisely real, nondegenerate eigenvalues
\cite{Smit87}.

All the properties expected to be important in QCD in 
four dimensions are present in QED2 and in two dimensions the full set of eigenvalues and 
eigenvectors can be computed on moderate size lattices~\cite{QED2d97,Smit87,SmitVink87}. 
Unlike in the continuum, the real eigenvalues do not occur at a specific 
value associated with the zero mass limit. In fact, even in a single 
configuration, there is no unique definition of the massless limit.
This limit can only be defined by an ensemble average. 
These fluctuations in the positions of real eigenvalues is a direct 
consequence of chiral symmetry breaking which occurs as an
artifact of Wilson-Dirac fermions.
Furthermore, this fluctuation in the positions of real eigenvalues will cause 
spurious poles in the fermion propagator for light fermion masses. This 
is the origin of the expectional configurations encountered in quenched
calculations.
Our detailed QED2 study of real eigenvalues near the mass zero continuum 
branch concluded the following~\cite{QED2d97}: (1) For fixed physical volume 
and light fermion mass, the problem with spurious poles (i.e. nearby real eigenvalues) 
is acute at strong coupling and decreases as $\beta$ increases; 
(2) The total number of real eigenvalues grows
with physical volume for fixed $\beta$; and (3) No decrease in the frequency of 
real eigenvalues in the physical light 
mass region is seen using $O(a)$ improved SW action.
%

%
\section{Exceptional Configurations in Lattice QCD}
%
%
 
In four dimensions, it is difficult to study the 
full eigenvalue spectrum\cite{Davies88}.  Fortunately for 
our purposes, we
only need to find the real eigenvalues near the mass zero 
continuum branch.

The zero modes appear as poles in the quark propagator
\begin{eqnarray}
&&(D-rW)f_i  =  \lambda_i f_i  \\
\lefteqn{~~~~~~S(x,y;\{U(A)\})_{AB}  = } \label{eq:eigenvals} \\
& & ~~~~~~~~~~~~~~~\sum_{eigenvalues} 
\frac{f_{iA}(x;U)\bar{g}_{iB}(y,U)}{\lambda_i + 1/2\kappa} \nonumber \\
&&m_q  =  1/2\kappa - 1/2\kappa_c \;
\end{eqnarray}
where $\kappa$ is the hopping parameter, with the critical value $\kappa_c$
determined from the ensemble ensemble average pion mass.
For modes with $\lambda_i < -1/2{\kappa_c}$, the propagator 
has poles corresponding to positive mass values.   The 
position of these poles can be established by studying any 
smooth projection of the fermion propagator as a function of 
$\kappa$.    
We will use the integrated 
pseudoscalar charge to probe for the shifted real eigenvalues
\begin{equation}
Q(\kappa) = \int d^4x<\bar{\psi}(x)\gamma_5\psi(x) >
\end{equation}
This charge is calculated using the same method used by 
Kuramashi et al.\cite{Kuramashi94}
to study hairpin diagrams and the ${\eta}'$ mass.  
The charge is a global quantity which samples the 
full lattice volume.    By computing its value for a range of 
kappa values we can search for poles in the fermion 
propagator.    Near a pole,  we should find
\begin{equation}
Q(\kappa) \rightarrow \frac{R}{1/\kappa-1/\kappa_{pole}}
\end{equation}
In the continuum, the residue R would be directly proportional 
to the global winding number of the gauge field
configuration\cite{SmitVink87}.

\begin{figure}
\vspace*{4.1cm}
\includegraphics{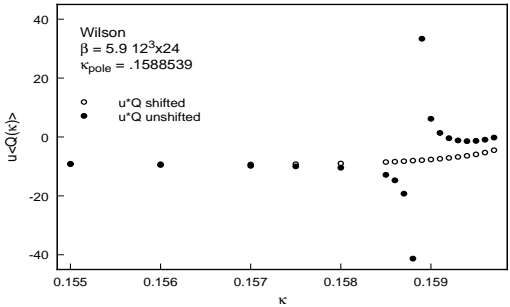}
\vspace*{-0.8cm}
\caption[]{$u <Q(\kappa)>$ versus $\kappa$ for one configuration.
$u = 2m_q$ and $<Q(\kappa)>$ 
is the average charge density. The unshifted charge
(solid dots) has $\kappa_{pole} = .1588539$. 
The MQA charge (open circles) 
is smooth (see Section 4).}
\label{fig:Q5poles}
\vspace*{-0.4cm}
\end{figure}

This procedure works well if the poles are in the visible region, 
$\kappa < \kappa_c$.   
There the pole positions and the relevant 
eigenvalues can be determined to great precision.    
An example of these scans is 
shown in Figure \ref{fig:Q5poles} for Wilson fermions.
%
%
%
%
The existence of isolated poles is obvious.  
The value of the integrated pseudoscalar charge can be computed, 
with no appreciable slowdown in convergence, for values of the hopping parameter
arbitrarily close to the pole position.

\begin{figure}
\vspace*{4.1cm}
\includegraphics{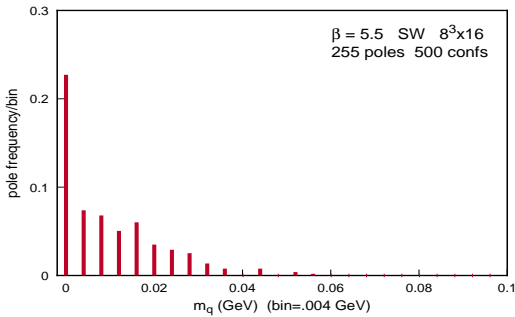}
\vspace*{-0.8cm}
\caption[]{The frequency of poles in the quark
propagator versus quark mass. The results use SW fermions on a $8^3\times 16$ lattice 
with $\beta = 5.5$. Analysis done by Jim Simone (unpublished).}
\label{fig:SWpoles}
\vspace*{-0.4cm}
\end{figure}

Scans for real poles in the light mass region 
have now been done for a wide variety of lattice couplings 
($\beta = 5.5, 5.7, 5.9, {\rm and} 6.1$) and 
volumes ($8^3\times 16, 12^3\times 24, 16^3\times 32$  and $ 24^3\times 48$)
with both unimproved and $O(a)$ improved Wilson fermions.
The frequency of poles versus $\beta$, volume, and quark mass
is in generally good agreement with expectations from QED2.
For example, at $\beta = 5.5$ on a 
$8^3\times 16$ lattice with $O(a)$ improved Wilson fermions 
($c_{SW}=1.67$),  255 poles were found in 500 configurations.
The distribution of poles for this case is shown in Figure \ref{fig:SWpoles}.
%
%
%
%
%

A detailed study was performed on a sample of 
50 gauge configurations on a $12^3\times 24$ lattice at $\beta = 5.9$ 
available from the ACPMAPS library \cite{f_b94}.    
We determined eigenvalues for
both the standard Wilson-Dirac action and a Clover 
action with a clover coefficient  of  $C_{SW} = 1.91$.
Six configurations were found with visible poles for each choice
of fermion action.    
These results are shown in Table I.  
Only one gauge configuration is in common between the 
two sets of visible poles.    

\begin{figure}
\vspace*{4.2cm}
\includegraphics{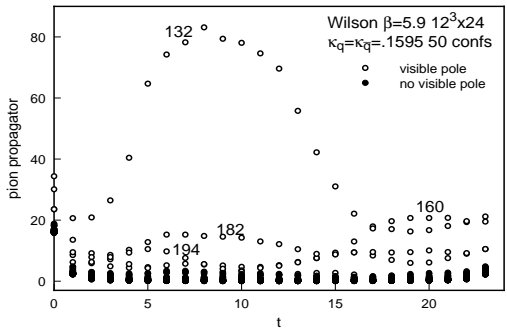}
\vspace*{-0.8cm}
\caption[]{Pion propagators for $\kappa =.1595$ 
showing all 50 configurations.  Visible poles (open circles) and 
all others (solid dots).}
\label{fig:pionset}
\vspace*{-0.4cm}
\end{figure}

The direct relationship between a nearby pole
in the quark propagator and an exceptional configuration at a given
quark mass is graphically demonstrated in Figure \ref{fig:pionset}.
There the time variation of all 50 pion propagators 
for $\kappa_q =\kappa_{\bar q}=0.1595$ is shown. 
The configurations already
identified in Table 1 as having a pole in the visible region
are indicated with open circles. The pion propagator is ``exceptional'' 
if and only if the configuration has a visible pole in Table 1.
Furthermore, the most exceptional configuration (132) has its  
visible pole at $\kappa=0.1594870$ closest to $\kappa=0.1595$.  

As the fermion action is varied, the real eigenvalues 
move.  Therefore, the visible 
poles, and the corresponding identification of exceptional 
configurations is a sensitive property of the fermion action 
and not a property unique to the particular gauge 
configuration.    For example, a small change in the 
clover coefficient may remove a visible pole for one 
configuration and add a visible pole for another 
configuration, completely changing identification of the 
exceptional configurations. Since only a collision of two real modes
allows the pairing required to move off the real axis,   
small changes in the parameters of the fermion action should 
not change the number of isolated real modes but only their visibility.
\begin{table}
\vspace*{-0.2cm}
\begin{center}
\caption{Visible poles for a set of 50 gauge configurations.}
\begin{tabular}{|c|c|c|}
\hline
\multicolumn{3}{|c|}{Wilson Action} \\
\hline
Conf. & $\kappa_{pole}$&        PS residue \\
\hline
114000 &0.1588539&      -2.1463 \\
132000 &0.1594870&      -2.4800 \\
148000 &0.1593216&      -3.6325 \\
160000 &0.1593803&      -2.8494 \\
182000 &0.1593803&      +2.5036 \\
194000 &0.1595557&      -5.3055 \\
\hline
\end{tabular}
\end{center}
\vspace*{-0.6cm}
\end{table}
%
%
%
%
%
\section{The Modified Quenched Approximation}
%
%
Since the effects of the fermion loop determinant are ignored 
in the quenched approximation, this  
approximation is very sensitive to singularities of the 
fermion propagator which may be encountered in particular
formulations of lattice fermions.   
In particular, the eigenvalue 
spectrum of Wilson-Dirac fermions generally contains a number 
of isolated real eigenvalues.   In the continuum limit, these 
eigenvalues are identified as zero modes and occur at precisely 
zero fermion mass.    In the Wilson-Dirac formulation, these eigenvalues 
shift due to the chiral symmetry breaking generated by 
the Wilson-Dirac action.    Some of these eigenvalues are shifted 
to positive mass which causes singular behavior for the 
fermion propagators computed for a mass near a shifted eigenvalue.    
In the quenched approximation, the shifts cause 
poles in the fermion propagators which are not  properly 
averaged.   This effect corresponds to similar situations in 
degenerate perturbation theory where small perturbative 
shifts can cause large effects due to small energy 
denominators.   In this case it is known that it is essential to 
expand around the exact eigenvalues and compensate the 
perturbation expansion with counter-terms in each order.    In 
the present case, a similar compensation must 
be made when using Wilson-Dirac fermions.   

We are now able to devise a procedure for correcting the 
fermion propagators for the artifact of the visible shifted poles. 
Consider the fermion 
propagator as a sum over the eigenvalues of the Wilson-Dirac 
operator (Eq.\ref{eq:eigenvals}). 
The shifted real eigenvalues cause poles at particular 
values of the hopping parameter.   The residue of the visible poles 
can be determined by computing the propagator for 
a range of kappa values 
close to the pole position and extracting the residue for the full 
propagator.  
With this residue we can define a modified quenched 
approximation by shifting the visible poles to 
$\kappa=\kappa_c$ and adding terms to 
compensate for this shift.
Although simply shifting visible poles to $\kappa_c$ would correct 
for the leading effects, it may distort an ensemble average. 
Since we are only able to identify poles with 
positive mass shifts, we have compensated a visible pole with 
one shifted to negative mass.   
These negative shifts do not 
generate singularities in the fermion propagators computed 
for positive mass values and are expected to cancel against 
poles with negative shifts generated by other configurations in 
the ensemble.   
With this procedure, we do not expect any 
large renormalization of $\kappa_c$ 
due to the shifting procedure.  

The full MQA fermion propagator may be simply computed 
by adding a term to the naive fermion propagator which 
incorporates a compensated shift of the visible poles.   The 
modified propagator is given by 
\begin{eqnarray}
\lefteqn{S^{MQA}(x,y;\{U(A)\})  \equiv S(x,y;\{U(A)\}) }\\
 & & ~~~~~~~~~~~~~~~~~~~+a_{pole}(\kappa)\times {\rm res}_{pole}(x,y) \nonumber
\end{eqnarray}
where
\begin{equation}
a_{pole}(\kappa) \equiv -\frac{1}{u-u_{pole}} 
+ \frac{2}{u} - \frac{1}{(u+u_{pole})}
\end{equation}
(At large mass the first two terms in the expansion in 
$1/u$ are not modified and terms linear in the shifts should 
average to zero), and  
\begin{eqnarray}
{\rm res}_{pole}(x,y)&\equiv&f_{pole}(x,U)\bar{g}_{pole}(y,U) \\
&=&\lim_{u \rightarrow u_{pole}}(u-u_{pole})*S(x,y,\{U\}) \nonumber
\end{eqnarray}
The residue of each pole is extracted by 
calculating the propagator at $u_{pole}-\Delta$ and $u_{pole}+\Delta$
with $\Delta\approx 10^{-5}$ where the pole position, 
$u_{pole}$, has been accurately determined from the 
integrated pseudoscalar charge measurement.   

It is important to note that the artifact is the appearance of visible 
poles at positive mass, not the existence of small or real 
eigenvalues.    It is only the visibility that we have corrected by 
appealing to the correct behavior in the continuum limit.    

%
\section{Applications}
%
%
MQA quark propagators 
may now  be used in place of the usual
quenched propagators.
The suppression of the large fluctuations
normally associated with the exceptional configurations
greatly reduces the errors associated with the 
propagation of light quarks. 
The most sensitive physical quantities are those associated with the 
chiral limit: the pion propagator,
the hairpin calculation for the $\eta'$ mass and topological
quantities.

We have determined fits to the pion mass, $m_{\pi}$, and 
coupling amplitudes, $|A_L|$ and $|A_S|$; using 
pion propagators computed with
the limited statistics of the 50 gauge configurations on a $12^3\times 24$ 
lattice at $\beta = 5.9$. 
We simultaneously fit 
the three pion two-point correlators LS(local-smeared), SL(smeared-local)
and SS(smeared-smeared) to a common pion mass.
Each two-point pion correlation
function, $G_{\pi\pi}(t)$ has the form:
\begin{equation}
 G_{\pi\pi}(t) = |A|^2 \frac{\cosh(m_{\pi}(T/2-t))}{\cosh(m_{\pi}T/2)} 
\label{eq:fit}
\end{equation}
%
%
\begin{figure}
\vspace*{4.6cm}
\includegraphics{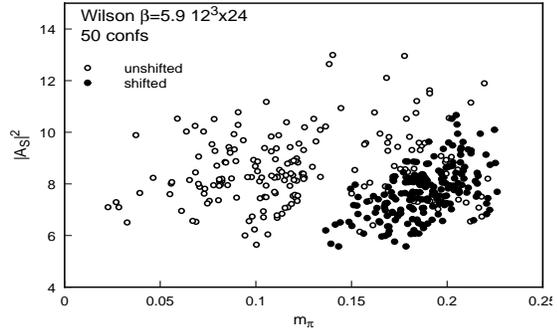}
\vspace*{-0.8cm}
\caption[]{Pion mass, $m_{\pi}$, and smeared coupling amplitude, 
$|A_S|$ for each of 200 bootstrap samples. 
Naive (open circles) and MQA (solid dots).}
\label{fig:Wboot}
\vspace*{-0.4cm}
\end{figure}
%
%

We determine fits using 200 bootstrap sets of 50 configurations,
In Figure \ref{fig:Wboot}, 
scatter plots compare the fluctuations and correlations for 
Wilson-Dirac fermions before and after using the MQA
analysis for $\kappa = 1.590$.    
It is clear that the MQA analysis greatly reduces
the fluctuations and produces a more tightly correlated
fit for the mass and decay constant.    
For the lighter quark masses, it is clear that the normal analysis 
would not be limited by statistics but by the frequency of
exceptional configurations associated with 
visible poles.    However, the MQA analysis
cures this problem and higher statistics would now greatly
improve the accuracy of computations with light mass
quarks.

\begin{figure}
\vspace*{7.5cm}
\includegraphics{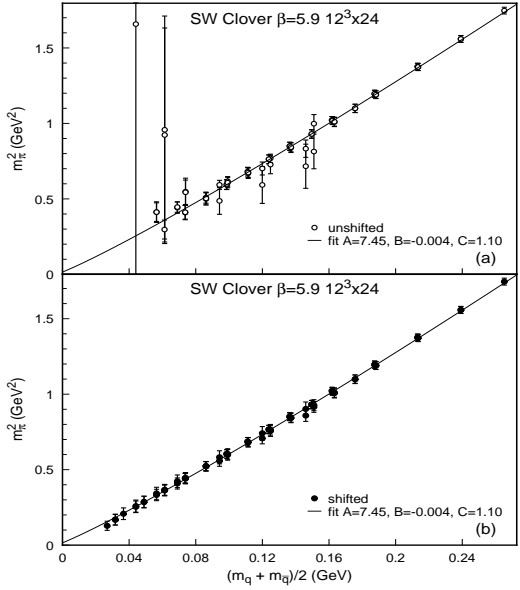}
\vspace*{-0.4cm}
\caption[]{Pion mass plots for Clover action:
(a) naive (open circles) and (b) MQA (dots).  A power law fit is also shown.}
\label{fig:pimassW1}
\vspace*{-0.4cm}
\end{figure}

In Figure \ref{fig:pimassW1}, we plot the square
of the measured values of the pion mass (for the Clover action)
against the average of the quark masses,
$m_l \equiv (m_{q1}+m_{q2})/2$, 
for the naive and MQA analysis, respectively.    
The large fluctuations in the naive analysis come when one or both
of the quarks are light.  
The masses determined from the 
MQA analysis seem to give a good fit to a nearly linear 
behavior. For general power law form, $ m_{\pi}^2 = A (m_l - B)^C$,
$A = m_{\pi}^2/m_l = 7.45$ and $C\approx 1.1$.   
%
%
%
%
For the naive Wilson action, the slope ($A=m_{\pi}^2/m_l = 4.30$)
is in good agreement with previous measurements\cite{f_b94}.  
The MQA fitted $\kappa_c=0.159725$ is very close 
to the value  $\kappa_c=0.15972$ obtained from the standard analysis.
The small shift in $\kappa_c$ reflects our use of 
a compensated shift for the visible poles and the linearity
observed in our mass fits.    

A detailed high statistics study on a variety of lattices of
these quenched chiral logs as well as 
the hairpin propagator and associated $\eta'$ mass~\cite{Thacker97}. 
is underway. Preliminary results
indicate that the MQA will allow a direct 
determination of the power law coefficient, 
$C = 1/(1+ \delta)$ in the relation
between $M_{\pi}^2$ and $m_l$ by using very light quark masses. 
%
%
\section{Conclusions}
%

The Wilson-Dirac operator has exact real
modes in its eigenvalue spectrum.   In the quenched approximation,
these real modes can generate unphysical poles in the valence
quark propagators for physical values of the quark mass.    These
poles can produce large lattice artifacts and are the source of
the exceptional configurations observed in attempts to directly
study QCD in the light quark limit.

The Modified Quenched Approximation identifies and replaces the
visible poles in the quark propagator by the proper zero mode
contribution, compensated to preserve  proper ensemble averages. 
All usual physical quantities can be computed with
only a modest overhead cost required to apply the full MQA analysis.
It allows stable quenched calculations with
very light quark masses and reduces the errors even in the
case of heavier quark mass.  Exceptional configurations are 
eliminated and hence errors can be meaningfully reduced by 
using larger statistical samples.

The usual O(a) improvement program does not remove the problem
of visible poles and exceptional configurations.
Indeed, we find the same size spread of
the real eigenvalues for both
Wilson-Dirac and Clover actions at the same lattice spacing.
This suggests the MQA analysis may be an essential ingredient
in realistic applications of improved actions on coarse
lattices.

Resolving the problem of exceptional configurations
removes a major obstacle to studying quenched
lattice QCD with Wilson fermions in the light quark limit.
%
%
%

\end{document}